\documentclass[12pt]{article}
\usepackage[top=1in, bottom=1in, right=1in, left=1in]{geometry} 
\usepackage{epsfig,fullpage,multirow,amsmath,amsfonts,natbib,latexsym, bm, parskip, setspace, arydshln ,titlesec}
\usepackage[affil-it]{authblk}
\usepackage{arydshln}
\usepackage[table]{xcolor}
\usepackage{lscape}
\usepackage{caption, subcaption}

\titlespacing*{\section}
{0pt}{8pt}{8pt}
\titlespacing*{\subsection}
{0pt}{8pt}{8pt}
\titlespacing*{\subsubsection}
{0pt}{0pt}{0pt}

\newcommand{\bfPhi}{{\bm \Phi}}
\newcommand{\bfphi}{{\bm \phi}}

\newcommand{\bfbeta}{{\bm \beta}}
\newcommand{\bfalpha}{{\bm \alpha}}
\newcommand{\bftheta}{{\bm \theta}}
\newcommand{\bfmu}{{\bm \mu}}
\newcommand{\bfeta}{{\bm \eta}}

\newcommand{\bfepsilon}{{\bm \epsilon}}

\newcommand{\bfSigma}{{\bm \Sigma}}

\newcommand{\bfgamma}{{\bm \gamma}}

\newcommand{\bftau}{{\bm \tau}}

\newcommand{\bfLambda}{{\bm \Lambda}}

\def\bfy{\mathbf{y}}

\def\bfB{\mathbf{B}}

\def\bfH{\mathbf{H}}

\def\bfX{\mathbf{X}}
\def\bfY{\mathbf{Y}}

\def\bfE{\mathbf{E}}

\def\bfI{\mathbf{I}}

\def\bfW{\mathbf{W}}

\def\bfzero{\mathbf{0}}

\def\bfzero{\mathbf{0}}

\makeatletter
\def\@maketitle{%
  \newpage
  \null
  \vskip 2em%
  \begin{center}%
  \let \footnote \thanks
    {\Large\bfseries \@title \par}%
    \vskip 1em%
    {\normalsize
      \lineskip .2em%
      \begin{tabular}[t]{c}%
        \@author
      \end{tabular}\par}%

    {\small \@date}%
  \end{center}%
 \vskip 1em
  \par
}

\title{A Bayesian change point model for spatio-temporal data}

 \author[1]{C.\ Berrett\thanks{cberrett@stat.byu.edu}}\affil[1]{Department of Statistics, Brigham Young University, Provo, UT, USA}
 \author[2]{B.\ Gurney\thanks{brianne.gurney@gmail.com}}\affil[2]{Department of Statistics, Brigham Young University, Provo, UT, USA}
  \author[3]{D.\ Arthur\thanks{david.b.arthur.90@gmail.com}}\affil[3]{Department of Statistics, Purdue University, West Lafayette, IN, USA}
   \author[4]{T. Moon\thanks{todd.moon@usu.edu}}\affil[4]{Department of Electrical Engineering, Utah State University, UT, USA}
 \author[5]{G.\ P.\ Williams \thanks{gus.p.williams@byu.edu}}\affil[5]{Department of Civil and Environmental Engineering, Brigham Young University, Provo, UT, USA}

\date{\today}

\begin{document}

\maketitle

\begin{abstract}
Urbanization can increase the recorded temperature representing the surrounding area. This phenomenon – a so-called urban heat island (UHI) – occurs at a local level at a point in time and has significant impacts on historical data analyses. This change is generally gradual, but can be abrupt, and occurs as construction or other urbanization changes the environment near a recording station.  We propose a methodology to examine if changes in temperature trends at a point in time exist at a given local level at various locations. Specifically, we propose a Bayesian change point model for spatio-temporally dependent data where we select the number of change points at each location using a “forwards” selection process using deviance information criteria (DIC). We then fit the selected model and examine the linear slopes across time to quantify changes in long-term temperature behavior. We show the utility of this model and method using a synthetic data set and observed temperature measurements from eight stations in Utah consisting of daily temperature data for 60 years.
 \end{abstract}

\section{Introduction}
When examining environmental data, researchers, governments, and other stakeholders are often interested in understanding changes in behavior over time and identifying relationships or causes. However, this can be a difficult task in practice.  The inherent spatial and temporal dependence found in environmental data can make it hard to discriminate between whether an observed ``change" is indeed a change or just part of the natural environmental behavior.  For example, urban heat islands (UHI's) are areas that see an increase in temperature compared to surrounding areas due to an increase in human activity or urbanization \citep{oke1973}.  While generally thought of as large metropolitan areas, UHI's can also be much smaller areas when, for example, farming land is replaced by an asphalt parking lot.  Because temperatures are often changing in natural (e.g., seasonal) and/or broad (e.g., regional or global) ways, identifying whether an observed change is natural, broad, or noise, or if it is a local change such as an UHI is precarious.  
Motivated by UHI's, we seek to propose a model that captures the natural and broad changes in temperature measurements across the state of Utah while simultaneously identifying and quantifying additional local changes.

\begin{figure}
    \centering
    \includegraphics[width=4in]{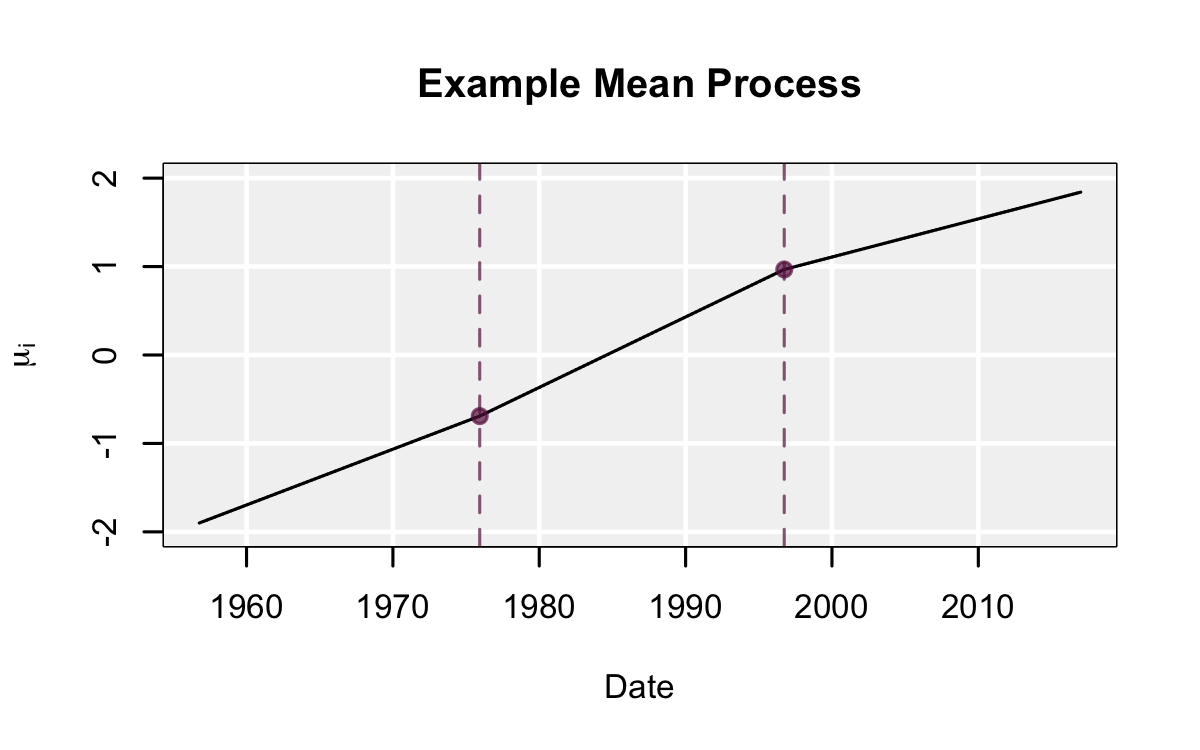}
    \caption{The black line shows an example of the mean process with two fixed change points (marked at the vertical dashed lines).}
    \label{fig:illustrate}
\end{figure}

UHI's are the result of local urbanization and therefore, future temperature measurements will not continue in the same manner \citep{Hoffman2012}.  An illustration of how an increasing temperature trend could change as urbanization occurs is shown in Figure \ref{fig:illustrate}.  At the first change point (e.g., as urbanization occurs near a recording station) denoted by the vertical dashed line between 1970 and 1980, the temperature starts increasing more rapidly for a period; at the second change point (e.g., as urbanization matures) denoted by the second vertical dashed line, the increasing temperature trend is lower.  Importantly, this change is occurring at the local level and not regionally.  Making use of regional data can help to identify temperature changes unique to a location.  

Increasing global and regional temperatures have been monitored closely from a variety of scientific perspectives for the last 150 years \citep{IPCC2019}.  However, many physical phenomena, such as UHI's, have effects that are eventually felt more widely but are generally experienced on a more local level \citep[see][for more information]{Balling1988, stabler2005, Hartz2006}.  
Changing temperatures are concerning to the general climate crisis, but local changes can impact health, energy, and other concerns at the local level \citep[see, e.g.,][]{Grimmond2007, Heaton2018}.  UHI's specifically were seen as a contributing factor to a heat wave that hit Chicago in July 1995, causing the death of approximately 525 people \citep{Changnon1996}.  In addition, UHI's contribute to an increase in demand for cooling systems and as such, power plants are under higher demand and release more carbon emissions \citep{Akbari2005}.  Furthermore, when urbanization occurs near a recording station, it can cause the data to exhibit trends that are not indicative of a larger area.  

Research has been done regarding factors contributing to this local effect.  For example, \cite{Weng2004} found that variations in land surface temperature due to vegetation in Indianapolis were responsible for existence and size of UHI's.  \cite{Saaroni2000} examined the spatial spread of UHI's through local temperatures at both street and roof levels in Tel Aviv.  Understanding the history of urbanization immediately surrounding a location is required to identify a UHI; thus, we seek to identify and quantify more general local behavior changes, if any, at locations across the state of Utah and if local records have been affected.  

We provide a framework for identifying general changes in the linear increase or decrease of temperature measurements.  We propose a Bayesian change point model as a tool to identify these changes.  Urbanization usually happens over a period of time, making resulting changes in temperature gradual and not abrupt.  As shown in Figure 1, this is exhibited as a chance in the temperautre trend over a longer period.  Thus, rather than observing a change in \emph{mean} temperature across different time periods, as is done in many change point models \citep[e.g.,][]{Fernhead2006, Benson2018}, we expect to observe a change in the \emph{slope} of the temperature across time.  These types of change points are not explored in the literature as much as changes in the mean.  \cite{Kim1989} explore maximum likelihood estimation of change points in slope, but the change occurs at a fixed time point and only accounts for one change point.  We assume changes occur instead at unknown periods of time, rather than fixed time points, and allow for multiple changes in the slope.  

The Bayesian framework is ideal for capturing this more gradual change in slope.  Rather than temperature changes occurring at a single point in time as in the illustration of Figure \ref{fig:illustrate}, a Bayesian model returns a posterior distribution -- or a range -- of times for the change.  The Bayesian change point model, however, is not without its challenges.  Specifically, short-term changes in temperature are common, making identifying the long-term changes of interest more difficult.  These short-term changes may be temporally or spatially correlated and represent the regional environment or climate trends.  Additionally, our data are over a long enough period of time that it's possible there may be more than one change in the temperature slope -- or none at all -- and Bayesian models allowing for these possibilities can be quite complicated, often requiring difficult-to-implement model fitting algorithms such as reversible jump Markov change Monte Carlo  \citep[MCMC][]{Green1995}.  Finally, the time of the change point and corresponding slopes will be highly correlated, making Markov change Monte Carlo (MCMC) convergence difficult \citep{Gregory2010}.

We tackle these challenges making use of multiple tools.  First, to aid in distinguishing between global/regional temperature patterns and more local temperature changes, we make use of spatio-temporal dependence by modeling measurements from multiple locations.  Second, we make use of variable selection techniques used in more typical regression analyses to select the number of change points at each location.  Third, we implement adaptive MCMC \citep{Haario2005} to accommodate the challenge of convergence. 

Bayesian spatio-temporal modeling has been used previously to determine change points.  \cite{Majumdar2005} sought to identify changes in temporal and spatial associations; however, the change points they investigated represented changes in the spatio-temporal dependence processes, not changes in mean or slope.  Further work has been contributed by \cite{Yu2008} who included temporal and spatial coefficients in their model to investigate the changes in assault rate decline over time.  They fit several models with different numbers of change points and different fixed times of each change point in parallel and compared models using the deviance information criteria \citep[DIC;][]{Spiegelhalter2002}.  We also use DIC, a measure of the likelihood of competing models, for model selection.  However, we make use of an automatic selection process for the number of change points, rather than a comparison of predetermined models.  

In our model, we allow the number of changes in slope to be different -- 0, 1, or 2 -- at each measurement location.  Thus, the total number of models to consider is $3^M$, where $M$ is the number of locations included in the model.  Even if $M$ is relatively small, the number of models to compare is large.  For eight locations there are 6,651 different models to consider.  Thus, to select the number of change points at each location, we propose a ``forward" selection, motivated by the forward selection used to select explanatory variables in multiple regression \citep[see, for example][]{Seber2003}.  This will allow us to find a good fitting model without exploring all possible models and will significantly ease computation of model exploration.  Notably, two change points at a single location for a period of 60 years (the length of the data we analyze) will enable us to capture long-term changes in slope.  If we were to allow more change points, we may risk overfitting the data by capturing shorter-term changes in slope that are more likely due to factors such as seasonal changes or temporary heat waves.  Additionally, using DIC as a criteria in our model selection process will help avoid overfitting concerns, as it includes a penalization for more complex models. 


To address issues of convergence, we take an adaptive MCMC approach.  Adaptive MCMC methods have been studied and shown to improve efficiency of the algorithm \citep{Haario2001, Roberts2009}.  The methodology can also improve convergence diagnostics, as demonstrated by \cite{Chauveau2002}.  In our analysis, we use the single component adaptive Metropolis \citep[SCAM;][]{Haario2005}.

The rest of the article is outlined as follows.  In Section \ref{sec:data} we provide information about the data used to illustrate the method.  Section \ref{sec:method} describes the general change-point model and model-fitting considerations.  The results of the proposed methodology and summary of the final fitted model are provided in Section \ref{sec:results}.  We finish with a discussion of considerations and future work in Section \ref{sec:discussion}.

\section{Data}\label{sec:data}

The data set we use to illustrate our methodology comes from the Utah State University Climate Center (USUCC) weather database \citep{USUCC2017}, which consists of temperature measurements from across locations in Utah for 123 years.  We use daily minimum temperature as minimum temperatures are more sensitive to urbanization than daily maximum temperatures.   

\begin{figure}
    \centering
    \includegraphics[width=5in]{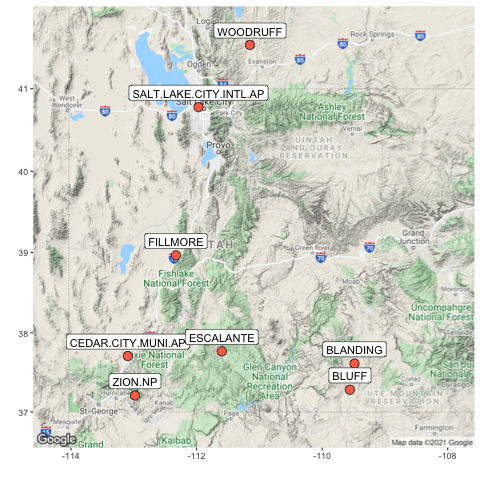}
    \caption{Map of Utah containing the eight locations (orange dots) of daily temperature readings used in this analysis.}
    \label{fig:map}
\end{figure}

We selected eight locations of daily minimum temperature (degrees Celsius) measurements from stations that covered the state of Utah and had less than five percent of data missing: Blanding, Bluff, Cedar City Municipal Airport, Escalante, Fillmore, Salt Lake City International Airport, Woodruff, and Zion National Park. These locations and their official station names are mapped in Figure \ref{fig:map} using the R package `ggmap' \citep{ggmap}. 

\begin{figure}
    \centering
    \includegraphics[width=6.5in]{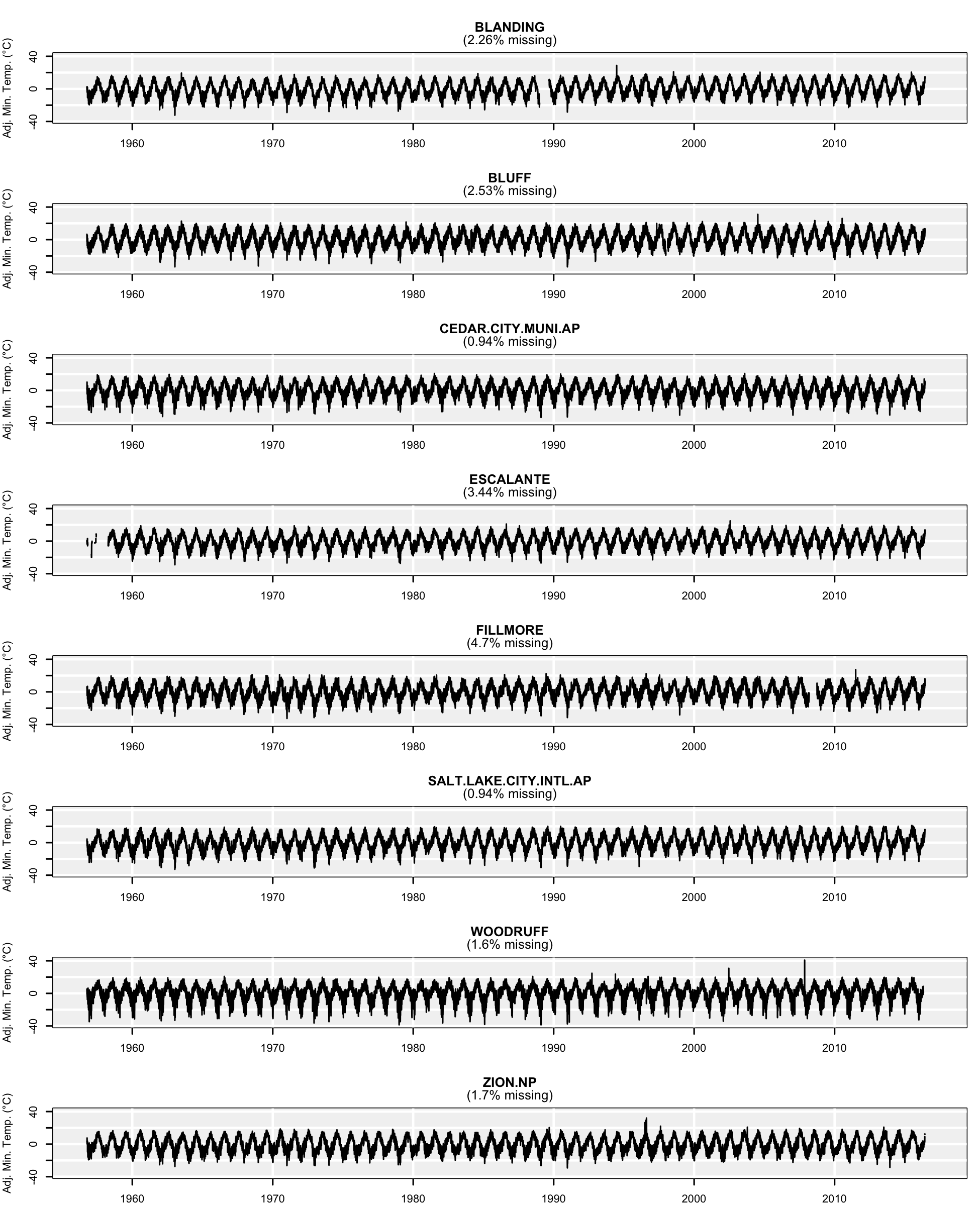}
    \caption{Zero-centered daily minimum temperature readings (Celsius) from each station.  Percent of days without a measurement for each location are indicated in the title.}
    \label{fig:dat}
\end{figure}

We collected over 60 years of daily temperature readings (October 8, 1956 -- December 31, 2016) and zero-centered the data at each location.  Figure \ref{fig:dat} shows these adjusted daily temperature measurements for each location.  The seasonality of minimum temperature is readily visible.  The linear increase or decrease in temperature, however, is not easily seen and any changes to this slope over time are even harder to see.  

\section{Methods}\label{sec:method}

We first outline the model for a given number of change points at each location in Section \ref{sec:model}.  Then in Section \ref{sec:forward} we outline our forward selection method for choosing the number of change points at each location.  In Section \ref{sec:consider} we provide some considerations for a few details of the model specification.   

\subsection{Model}\label{sec:model}

Let $y_{it}$ be the (zero-centered) temperature measurement at location $i$ and time $t$.  We model
\begin{equation}
y_{it} = \mu_{it} + \phi_{it} + \epsilon_{it},\label{eq:smmod}
\end{equation}
where $\mu_{it}$ represents the mean process, $\phi_{it}$ the spatio-temporal process, and $\epsilon_{it}$ residual noise.  We assume $\epsilon_{it} \stackrel{iid}{\sim} \mathcal{N}(0, \sigma^2_\epsilon)$, where $\sigma^2_\epsilon$ is the variance.  

Using vector notation, we can write (\ref{eq:smmod}) for all locations and times under consideration.  Let $\bfy_i = (y_{i1}, \dots, y_{iN})'$ be an $N \times 1$ vector of temperature measurements at location $i$, where $N$ represents the total number of days being considered and $\bfY = (\bfy_1', \dots, \bfy_M')'$ be the $NM \times 1$ vector of observations for all $M$ locations.  Then, 
\begin{equation}
    \bfY = \bfmu + \bfPhi + \bfepsilon, \label{eq:bigmod}
\end{equation}
where $\bfmu$, $\bfPhi$, and $\bfepsilon$ represent the corresponding $NM \times 1$ vectors of the mean process, spatio-temporal process, and noise, respectively.  

The mean process, $\bfmu$, accounts for a changing linear trend in time -- including any relevant change points -- at each location.  Using standard linear model notation, $\bfmu$ can be written in the form $\bfX\bfbeta$.  Consider the mean process for location $i$ at time $t$.  We represent the mean process for a fixed number of change points $q_i$, as
\begin{align}
    \mu_{it} = & \, \beta_{0i} + \beta_{1i}\,t^*\,I(t \le \tau_{1i}) + (\beta_{2i} - \beta_{1i})\,\tau_{1i}^*\,I(t > \tau_{1i}) + \beta_{2i}\,t^*\,I(\tau_{1i} < t \le \tau_{2i}) + \cdots + \label{eq:meanproc}\\ \nonumber
    & \beta_{q_i i}\,t^*\,I(\tau_{q_i-1,i} < t \le \tau_{q_ii}) + (\beta_{q_ii} - \beta_{q_i+1,i})\, \tau_{qi}^*\, I(t > \tau_{q_ii}) + \beta_{q_i+1,i}\, t^*\, I(t > \tau_{q_ii}), 
\end{align}
where $t^*$ represents the centered and scaled day, $\beta_{qi}$ represents the coefficient corresponding to the $q$th change point, and $\tau_{qi}$ is the day of the $q$th changepoint.  To help facilitate the adaptive MCMC algorithm, we assume that $\tau_{qi}$ is continuous; for use in (\ref{eq:meanproc}), $\tau^*_{qi}$ is $\tau_{qi}$ rounded up to the nearest day and then scaled to match the scaling of $t^*$.  In our model, we scale $t^*$ so that it is between $-1$ and $1$.  

It can be helpful to re-arrange the coefficients in Equation (\ref{eq:meanproc}) so that, for $q_i=2$, $\mu_{it}$ can be written in the following form,
\begin{align}
\mu_{it} = &\, \beta_{0i} + \beta_{1i}\, \left[ (t^*\, I(t \le \tau_{1i}) + \tau_{1i}^*\,I(t > \tau^*_{1i})\right] \nonumber\\ 
&\,+ \beta_{2i}\,\left[\max\{0, t^* - \tau^*_{1i}\} I(t \le \tau_{2i}) + (\tau^*_{2i} - \tau^*_{1i} ) I (t > \tau_{2i})\right] \nonumber \\ 
\,&+\beta_{3i}\, \left[\max\{0, t^* - \tau^*_{2i}\} I(t \le \tau_{2i}) + (t^* - \tau^*_{2i}) I(t > \tau_{2i})\right]. \label{eq:meanproc2}
\end{align}
Figure \ref{fig:illustrate} shows how the different coefficients contribute to the mean function.  In this figure, we fix $\beta_{0i} = 0$, $\beta_{1i} = 1.9$,  $\beta_{2i} = 2.4$,   $\beta_{3i} = 1.3$, $\tau_{1i} = 7000$ (day that corresponds to December 7, 1975), and $\tau_{2i} = 14600$ (day that corresponds to September 27, 1996).  By parameterizing the mean process this way, the coefficients represent the slope between corresponding time periods.  

Extending Equation (\ref{eq:meanproc2}) to matrix notation, let $\bfbeta_i = (\beta_{0i}, \beta_{1i}, \dots, \beta_{q_i+1,i})'$ be the $(q_i + 2) \times 1$ vector of coefficients and $\bfX_i$ be the corresponding $N \times (q_i + 2)$ matrix.  Then, $\bfmu_i = \bfX_i\bfbeta_i$ for each location.  The mean process for all locations can then be represented using an $NM \times (2M + Q)$ block diagonal matrix $\bfX$, where $Q = \sum_{i=1}^M q_i$ is the total number of change points in the model and the $i$th block diagonal of the matrix $\bfX$ contains the matrix $\bfX_i$.  The corresponding $(2M + Q) \times 1$ vector of coefficients is $\bfbeta = (\bfbeta_1', \dots, \bfbeta_N')'$.  Thus, 
\begin{equation}
    \bfmu = \bfX\bfbeta. \label{eq:meanproc3}
\end{equation}

To account for spatio-temporal dependencies between temperature readings, we choose to use basis functions.  Using basis functions to model spatio-temporal dependence reduces the dimensions and increases the ease of computation \citep[see, e.g.,][]{Cressie2008}. Let $\bfphi_t = (\phi_{1t}, \dots, \phi_{Mt})'$ be the $M \times 1$ vector representing the spatial process at time $t$.  We model this process such that, 
\begin{equation}
    \bfphi_t = \bfB\bfgamma_t,
\end{equation}
where $\bfB$ is an $M \times K$ matrix of spatial basis functions, $\bfgamma_t = (\gamma_{1t}, \dots, \gamma_{Kt})'$ is a $K \times 1$ vector of coefficients, and $K$ is the choice for number of spatial bases.  

To account for temporal dependence, we model $\bfgamma_t$ using temporal bases.  Since the number of time points -- and therefore the number of $\bfgamma_t$'s -- is very large, a basis function expansion to reduce the dimensions is computationally necessary.  Let $\bfgamma_k = (\gamma_{k1}, \dots, \gamma_{kN})'$ be the $N \times 1$ vector representing the coefficients for the $k$th spatial basis.  Then, 
\begin{equation}
    \bfgamma_k = \bfH \bfalpha_k,
\end{equation}
where $\bfH$ is the $N \times L$ matrix of temporal bases, and $\bfalpha_k$ is an $L \times 1$ vector of coefficients.  

There are many possible spatial and temporal basis functions to use for $\bfB$ \citep[e.g.,][]{Calder2002, Katzfuss2012} and $\bfH$ \citep[e.g.,][]{Zumer2008, Furgale2012, Novosad2016}; our choice for spatial and temporal basis functions will be described in Section \ref{sec:consider}.

The spatial process, $\bfPhi$, can then be written using $\bfB$ and $\bfH$.  Let $\bfB^* = (\bfB \otimes \bfI_N)$ and $\bfH^* = (\bfI_K \otimes \bfH)$, where $\bfI_N$ is the $N \times N$ identity matrix.  Additionally, let $\bfalpha = (\bfalpha_1', \dots, \bfalpha_K')'$ be the vector of all spatio-temporal coefficients.  Then, 
\begin{equation}
    \bfPhi = \bfB^*\bfH^*\bfalpha. \label{eq:spatproc}
\end{equation}
Note that $\bfB^*\bfH^* \equiv (\bfB \otimes \bfH)$.  

Making use of Equations (\ref{eq:meanproc3}) and (\ref{eq:spatproc}), Equation (\ref{eq:bigmod}) can be written as
\begin{equation}
    \bfY = \bfX\bfbeta + \bfB^*\bfH^*\bfalpha + \bfepsilon.
\end{equation}
Thus, the likelihood of $\bfY$ is
\begin{equation}
    \bfY | \bfX, \bfbeta, \bfalpha, \sigma^2_\epsilon \sim \mathcal{N}(\bfX\bfbeta + \bfB^*\bfH^*\bfalpha, \sigma^2\,\bfI_{NM}).
\end{equation}

For all parameters other than the change points, we use conjugate prior distributions, specifically, \begin{align*}
    \bfbeta &\sim \mathcal{N}(\bfzero_\bfbeta, \bfSigma_\bfbeta),\\
    \bfalpha &\sim \mathcal{N}(\bfzero_\bfbeta, \bfSigma_\bfalpha),\\
    \sigma^2_\epsilon &\sim \mbox{InvGamma}(a_{\sigma^2_\epsilon}, b_{\sigma^2_\epsilon}),
\end{align*}
where $\bfzero_\bfbeta$ and $\bfzero_\bfalpha$ represent the vectors of zeros with the same length as $\bfbeta$ and $\bfalpha$, respectively, $\bfSigma_\bfbeta$ and $\bfSigma_\bfalpha$ represent the prior covariance matrices for $\bfbeta$ and $\bfalpha$, and $a_{\sigma^2_\epsilon}$ and  $b_{\sigma^2_\epsilon}$ represent the shape and scale parameters so that the prior expected value of $\sigma^2_\epsilon$ is $b_{\sigma^2_\epsilon}/(a_{\sigma^2_\epsilon}-1)$.

The prior distributions for each change point, $\tau_{qi}$, is dependent on the number of change points at location $i$.  If location $i$ has one change point, then, 
\begin{equation}
    \tau_{1i} \sim \mbox{Unif}(b_i, N-b_i),
\end{equation}
where $b_i$ is a bound we choose to avoid change points too close to the beginning or end of the time period.  If location $i$ has two change points, then,
\begin{align}
    \tau_{1i} & \sim \mbox{Unif}(b_i, N - b_i),\\
    \tau_{2i}|\tau_{1i} & \sim \mbox{Unif}(\tau_{1i} + \delta_{i}, N-b_i),
\end{align}
where $\delta_i$ is a bound that separates the two change points.  Without the bound, posterior distributions for both change points may be too near each other and not provide distinct change points.  Thus, the bound allows for some separation.  These priors can be further expanded to include locations with more than two change points, adding in a new $\delta$ parameter to separate subsequent change points.  

Since the prior distributions for $\bfbeta$, $\bfalpha$, and $\sigma^2_\epsilon$ are conjugate with the likelihood, the full conditional distributions can be easily derived.  These are, 
\begin{align*}
    \bfbeta|\bfalpha, \sigma^2_\epsilon, \bftau, \bfY & \sim \mathcal{N}\left(\frac{1}{\sigma^2_\epsilon}\bfSigma^*_\bfbeta \bfX'(\bfY - \bfB^*\bfH^*\bfalpha), \bfSigma^*_\bfbeta\right),\\
    \bfalpha | \bfbeta, \sigma^2_\epsilon, \bftau, \bfY & \sim \mathcal{N}\left( \frac{1}{\sigma^2_\epsilon}\bfSigma^*_\bfalpha(\bfB^*\bfH^*)'(\bfY - \bfX\bfbeta), \bfSigma^*_\bfalpha\right),\\
    \sigma^2_\epsilon|\bfbeta, \bfalpha, \bftau, \bfY & \sim \mbox{InvGamma}\left(a_{\sigma^2_\epsilon} + 0.5\,NM, b_{\sigma^2_\epsilon} + 0.5\, \bfE'\bfE \right),
\end{align*}
where $\bftau$ represents a vector with all change points across all locations and 
\begin{align*}
    \bfSigma^*_\bfbeta &= \left( \frac{1}{\sigma^2_\epsilon}\bfX'\bfX + \bfSigma_\bfbeta\right)^{-1}, \\
    \bfSigma^*_\bfalpha & = \left(\frac{1}{\sigma^2_\epsilon}(\bfB^*\bfH^*)'(\bfB^*\bfH^*) + \bfSigma_\bfalpha\right)^{-1},\\
    \bfE &= (\bfY - \bfX\bfbeta - \bfB^*\bfH^*\bfalpha).
\end{align*}
There is not a conjugate prior for the $\tau_{qi}$'s.  Thus, we use an adaptive Metropolis random walk algorithm, making use of the SCAM algorithm \citep{Haario2005}.

\subsection{Model Selection}\label{sec:forward}

The columns of the $\bfX$ matrix depend on the change points in the model -- both how many change points and the time of each change point.  The model requires the number of change points to be fixed and then estimates the times of the change points.  Fixing the number of change points in the model is practical since the time of the change point can only be determined if a change point exists.  Thus, we propose a method for choosing the number of change points at each location.

This analysis is motivated by identifying long-term changes in temperature behavior, therefore, we limit the number of possible change points at any given location to be 0, 1, or 2.  Of course this method could be expanded to have more change points for other applications.  As mentioned before, with three different possible number of change points at each location, the total number of possible models to compare would be $3^M$, which increases exponentially as the number of locations increases.  Instead of comparing all models, to reduce computational burden, we propose a ``forward" change point selection process based on DIC.  We use DIC because of its ability to account for model complexity and its efficacy in parallel model selection, as demonstrated by \cite{Congdon2005}.  

\cite{Spiegelhalter2002} defines DIC to be
\begin{equation}
    DIC = p_D + \bar{D},
\end{equation}
where $p_D$ represents the number of effective parameters int he model and $\bar{D}$ represents the mean deviance.  $\bar{D} = (1/N_D) \sum_{n=1}^{N_D}D_n$, where $N_D$ represents the number of deviances calculated from the samples of the posterior distribution and $D_n$ represents the deviation calculated at the $n$th posterior sample.  Specifically, 
\begin{equation}
    D_n \equiv D(\bftheta_n) = -2 \log f(\bfy|\bftheta_n),\label{eq:deviance}
\end{equation}
where $\bftheta_n$ is the $n$th draw of the vector of posterior samples of $\bftheta$ and $f(\cdot)$ is the likelihood defined in Equation (\ref{eq:bigmod}).  Furthermore, $p_D = \bar{D} - D(\bar{\bftheta})$, where $D(\bar{\bftheta})$ is the deviance in Equation (\ref{eq:deviance}) evaluated at the posterior mean of $\bftheta$.  For computing the DIC using Equation (\ref{eq:bigmod}), $\bftheta = (\bfmu', \bfPhi', \sigma^2_\epsilon)'$, using equations (\ref{eq:meanproc3}) and (\ref{eq:spatproc}) to compute the posterior sampled values of $\bfmu$ and $\bfPhi$.

We use this definition of DIC as the criteria for selecting models in our forward selection process.  As with forward selection in regression, the idea is to slowly add change points at each location, fit the model for that combination of change points, and compare the resulting DIC.  The process will ideally find a model that minimizes, or nearly minimizes, DIC and is thus a good fit for our data.  

We additionally choose models based on convergence, as some models may not converge.  For example, if a model has too many change points, the posterior distribution for the time of the change point may cover the entire time period of the analysis, making convergence difficult.  We assume that a model that converges is more likely to contain the correct number of change points than a model that does not converge.  We explore this in our simulation study described in Section \ref{sec:simulation}. 

To assess convergence automatically within the proposed forward selection process, we examine the effective sample size (ESS) of all parameters and the posterior standard deviation of the change points.  We use the effective sample size as defined by \cite{Gong2016} and compute using the ``mcmcse" R package \citep{Flegal2020}.  ESS is smaller for samples with strong autocorrelation.  As with many Bayesian models, we expect autocorrelation to be high.  Our simulation study indicated that the ``true" model never had a parameter with ESS less than 1.5\% of the number of saved posterior draws.  Thus, we select from models where the ESS for all parameters are above 1\% of the saved posterior draws.  We also select from models where the posterior standard deviation of all the change points are less than 2500 days. 

The forward selection process has the following main steps: 
\begin{enumerate}
    \item Begin with the model where no change points are present in any location.
    \item Add one change  point to each location, creating $M$ more models to consider.  An example of this where $M = 8$ is shown in Table \ref{tab:forward}.
    \item Obtain samples from the posterior distributions of the model parameters by fitting each model via MCMC.  
    \item Compute and compare DIC values for each model.  From the models that meet the ESS and posterior standard deviation criteria, select the model with the smallest DIC.  (If none of the models converge, either stop the selection process, or choose the model with the smallest DIC.)
    \item Using the chosen model, add one change point to each of the locations, creating $M$ new models, similar to Step 2.  If there are more than two change points at any location for any model, do not fit that model.  
    \item Repeat steps 3--5 until either no models converge or no more change points can be added (i.e., all locations have two change points).  
    \item Compile all models generated by the process and locate, from the converged models, the model with the smallest DIC.  This will be the model chosen for inference.  
\end{enumerate}

A backwards or step-wise selection process could also be considered.  

\begin{table}
    \centering
        \caption{The number of change points at each location for the first set of models fit in the forward selection process.}
    \label{tab:forward}
    \begin{tabular}{c|cccccccc}
    \hline
    & \multicolumn{8}{c}{Location}\\
        Model & A & B & C& D& E &F &G& H \\\hline\hline
         1& 0 & 0 & 0 & 0 &0&0&0&0\\
          2& 1 & 0 & 0 & 0 &0&0&0&0\\
           3& 0 & 1 & 0 & 0 &0&0&0&0\\
            4& 0 & 0 & 1 & 0 &0&0&0&0\\
             5& 0 & 0 & 0 & 1 &0&0&0&0\\
              6& 0 & 0 & 0 & 0 &1&0&0&0\\
               7& 0 & 0 & 0 & 0 &0&1&0&0\\
                8& 0 & 0 & 0 & 0 &0&0&1&0\\
                 9& 0 & 0 & 0 & 0 &0&0&0&1\\\hline
    \end{tabular}
\end{table}

\subsection{Model Considerations}\label{sec:consider}

Three additional considerations are necessary for defining the model: choosing the spatial and temporal bases, the number of bases to use ($K$ and $L$), and the boundary values for the change point prior distributions.  

The number of bases used for the spatio-temporal dependence has some trade-offs.  In general, fewer bases make a smoother spatio-temporal process and is easier to fit computationally.  In contrast, more bases are not as smooth and will be more sensitive to short-term changes in temperature, thus possibly overfitting the data.  Examples of such basis function expansions are explored by, for example, \cite{Higdon2002, Lee2011, Sloan2017}.  Additionally, because the change points are being modeled in the mean process, care must be taken to ensure that the bases and $\bfX$ are not collinear.  

We represent the temporal dependence using the Fourier basis function expansion.  These basis functions are built from the Fourier series and Fourier transform \citep[see, e.g.,][]{Stein1971}.  This representation of the basis function has been found to compare favorably to other common basis function expansions \citep{Konidaris2008}
 and is thus applicable to represent the periodic temporal dependence.  The $\ell$th Fourier basis function at time $t$, $\xi_\ell(t)$, is defined as
 \begin{equation}
     \xi_\ell(t) = \begin{cases}
     0, & \mbox{if } \ell = 0\\
     \cos\left(\frac{\ell + 1}{A}\pi t\right), & \mbox{if } \ell > 0 \mbox{ and } \ell \mbox{ is odd}\\
     \sin\left(\frac{\ell}{A}\pi t\right), & \mbox{if } \ell > 0 \mbox{ and } \ell \mbox{ is even}
     \end{cases},
 \end{equation}
where $A$ represents the estimated period for the time series, in our case $A = 365.25$ \citep{Konidaris2008}. Thus, in our analysis, the $\ell$th column and $t$th row is equal to $\xi_\ell(t)$.  

We model spatial dependence using empirical orthogonal functions \citep[EOF's;][]{Navarra2010}.  Because the data follow very similar seasonal behavior, we compute the EOF's based on the correlation between locations after subtracting the fitted linear and seasonal trend from each location's measurements.  Let $\bfeta_i$ be the $N \times 1$ vector of residuals for the $i$th location, such that,
\begin{equation}
    \bfeta_i = \bfy_i - \bfX_{i0}\hat{\bfbeta}_i - \bfH\hat{\bfalpha_i},
\end{equation}
where $\bfX_{i0}$ is the $\bfX_i$ matrix with zero change points, $\hat{\bfbeta}_i$ and $\hat{\bfalpha}_i$  are the maximum likelihood estimates of the linear model fit to the $i$th location of temperature measurements.  Let $\bfSigma_\bfeta$ be the $M \times M$ correlation matrix with the $ij$th element equal to the correlation between $\bfeta_i$ and $\bfeta_j$.  The eigendecomposition of the matrix $\bfSigma_\bfeta$ is 
\begin{equation}
    \bfSigma_\bfeta = \bfW \bfLambda \bfW^{-1},
\end{equation}
where $\bfW$ is an $M \times M$ matrix consisting of the eigen vectors associated with each eigen value found along the diagonal of the $M \times M$ matrix $\bfLambda$.  The EOF's are the columns of $\bfW$ and the matrix $\bfB$ contains the first $K$ of these vectors.  

We use the mean-squared error (MSE) to select the number of temporal bases.  For different values of $L$ (2, 4, 6, 8, and 10), we compute the MSE $\sum_{i=1}^M \bfeta_i'\bfeta_i$.  At a certain point, adding more bases only minimally changes the MSE.  We choose the number of bases that reduces the MSE from the prior number of bases by less than 1\%.  In our data analysis, $L = 8$.  

The number of spatial bases is chosen based on the eigenvalues of $\bfSigma_\bfeta$, using the eigenvectors corresponding to the largest eigenvalues.  In our data analysis, $K = 3$.  Images of the values of the bases for each location of the data analysis are provided in the Supplementary Document.

A final consideration for this model is the boundary points for the change points.  We consider boundary points where change points are too close to each other or at the edges of the time period.  Since temperature changes are gradual over time, we assume there will not be meaningful change points within a small time period of each other.  Any change points that exist too closely together should essentially be swept up into the posterior distribution of one change point, since they account for the same gradual change in temperature. 

In some sense, the boundaries can help with convergence since they help to avoid the issue of overlapping posterior distributions for more than one change point.  However, these bounds can also add convergences issues.  For example, say the upper boundary for possible change points at a location is $t=20000$.  If a true change point were to exist at $t=19999$, getting draws at this value would be difficult, since all draws above $t=20000$ would be rejected.  Thus, the few accepted values will have a small ESS and not converge.  However, this issue could also arise when forcing more change points than are necessary.  If the model attempts to fit two change points to a location where only one (or zero) exists, the second will not converge.  The algorithm will continue to search for such a change point until it reaches the upper boundary, at which point it may accept some values.  Thus, convergence issues in models with both change points near the boundaries may or may not be due to the wrong choice of number of change points.  Thus, the boundary points may be chosen to exclude identifying change points at or near the boundary that are not of interest to the researcher.  

\section{Data Analysis}\label{sec:results}

We first use a synthetic (simulated) data set to examine the efficacy of the proposed methodology.  Specifically, we perform a simulation study to: examine if the proposed forward selection process identifies the data-generating model, determine the number of posterior samples needed for the DIC to differentiate between the models, and select cut-offs for the ESS and posterior standard deviation convergence criteria.   We then apply the proposed methodology to temperature measurements from eight locations across Utah, as described in Section \ref{sec:data}.  We examine the model selected and take a close look at the inference from the selected models and its implications for changing temperatures at each location.

\subsection{Synthetic Data}\label{sec:simulation}

To determine the efficacy of our model and method, we create a synthetic data set of the same size as the temperature data, specifically 22,000 observations for eight locations.  The ``true" number of change points are shown in the highlighted row of Table \ref{tab:simfwd}.  The temporal and spatial bases were defined using the observed data as described in Section \ref{sec:consider} with a choice of four temporal bases and one spatial basis, as was selected by \cite{Arthur2018}.  The parameter values used to create the synthetic data are provided in the Supplementary Document.  

We ran the MCMC algorithm for 122,000 iterations, discarded 2,000 as burn-in, and thinned by five for memory purposes.  This left 24,000 saved draws from the posterior distributions of each parameter.  Prior parameters are $\bfSigma_\bfbeta = 10,000\,\bfI$, $\bfSigma_\bfalpha = 10,000\,\bfI$, $a_{\sigma^2_\epsilon} = 10$,  $b_{\sigma^2_\epsilon} = 1$, and $b_i = \delta_i = 2000$ for all locations.  We make use of the ``parallel" package in R \citep{R2021} to fit the models in parallel.

Table \ref{tab:simfwd} shows the change points for each model chosen at each step of the forward selection process.  Each model was chosen because it had the smallest DIC value compared to the other models fit in that step.  Thus, the model with the smallest DIC in the table is also the model with the smallest DIC of all 82 models fit.  Happily, the model with the smallest DIC had the same number of change points as the true number used to create the data.  Thus, the forward selection process was effective at choosing the correct model for the synthetic data set.  We repeat the selection process with different random seeds and the results are the same -- the method repeatedly selects the model with the true number of change points.

\begin{table}[]
    \centering
        \caption{The change points for each model chosen at each step of the forward selection process, with the model with the smallest DIC (and the true generating model) highlighted in gray.}
    \label{tab:simfwd}
    \begin{tabular}{c|cccccccc|cr}
    \hline
    & \multicolumn{8}{c|}{Location} & \\
Step & A & B& C & D & E & F & G& H & DIC & Min ESS\\\hline \hline
1 &   0 & 1 & 0 & 0 & 0 & 0 & 0 & 0   & 1068613.7 & 2114.72\\
2 &   0 & 1 & 0 & 1 & 0 & 0 & 0 & 0   & 1068525.5 & 528.61 \\
3 &   0 & 1 & 0 & 1 & 1 & 0 & 0 & 0   & 1068459.4 & 516.37 \\
4 &   0 & 1 & 0 & 1 & 2 & 0 & 0 & 0   & 1068360.3 & 656.21 \\
\rowcolor{lightgray}5 &   0 & 1 & 1 & 1 & 2 & 0 & 0 & 0   & 1068334.7 & 458.84 \\
6 &   0 & 1 & 2 & 1 & 2 & 0 & 0 & 0   & 1068335.6 & 356.64 \\
7 &   0 & 1 & 2 & 1 & 2 & 0 & 1 & 0   & 1068338.3 & 346.46\\
8 &   1 & 1 & 2 & 1 & 2 & 0 & 1 & 0   & 1068338.5 & 328.7\\
9 &   1 & 1 & 2 & 1 & 2 & 0 & 2 & 0   & 1068339.8 & 309.97\\
10 &  2 & 1 & 2 & 1 & 2 & 0 & 2 & 0   & 1068341.6 & 355.19 \\
11 &  2 & 1 & 2 & 1 & 2 & 0 & 2 & 1   & 1068343.2 & 412.72 \\
12 &  2 & 1 & 2 & 1 & 2 & 1 & 2 & 1   & 1068345.1 & 425.44 \\
13 &  2 & 1 & 2 & 1 & 2 & 2 & 2 & 1   & 1068347.2 & 467.14\\
14 &  2 & 1 & 2 & 1 & 2 & 2 & 2 & 2   &
1068349.1 & 321.73 \\
15 &  2 & 2 & 2 & 1 & 2 & 2 & 2 & 2   & 1068350.5 & 36.99 \\
16 &  2 & 2 & 2 & 2 & 2 & 2 & 2 & 2   & 1068352.0 &  9.36\\\hline
    \end{tabular}

\end{table}

Some of the DIC values in Table \ref{tab:simfwd} are very close to one another.  Because we use MCMC to approximate the posterior distributions, the computed values of DIC are also approximate. Thus, there will be some possibility that the DIC value from a wrong model will be smaller than the DIC from the true model simply due to Monte Carlo error. We therefore use the synthetic data set to examine possible DIC values for the true model.  We compare this distribution of computed DIC values to the computed values from two wrong models: one similar to the truth and one completely different from the truth.  For the method to correctly distinguish between the true and wrong models, the range of computed DIC values from the true model should be distinct from and lower than the DIC values from the wrong models.  

Table \ref{tab:simmods} shows the number of change points fit at each location for the three different models.  We designed the similar model to have only one location with a different number of change points from the truth.  We designed the wrong model to have a completely different number of change points at every location.  

\begin{table}
    \centering
        \caption{Number of change points at each constructed location for three models fit to the synthetic data to compare the DIC.}
    \label{tab:simmods}
    \begin{tabular}{c|cccccccc}
    \hline
    & \multicolumn{8}{c}{Location}\\
        Model &  A & B& C& D& E& F& G& H \\ \hline \hline
        True Model &  0 & 1 & 1 & 1 & 2& 0 & 0 & 0 \\
        Similar Model & 0 & 2 &1 &1 & 2 & 0 & 0  &0\\
        Wrong Model & 2 & 2& 0 &2&0&1&2&1\\\hline
    \end{tabular}
\end{table}

Each model listed in Table \ref{tab:simmods} is fit 20 separate times using different random seeds.  We compare the DIC values from the 20 fitted models for different number of saved posterior draws.  After 2,000 burn-in and thinning by every fifth draw, we examine the DIC values for 2,000, 8,000, 16,000, and 20,000 saved posterior draws.  

We use an overlapping index to measure the distinctness of the fitted DIC values for each model.  \cite{Pastore2019} propose  an overlapping index that can be thought of as a measure of misclassification, where 0 indicates no overlap (the distributions are distinct), and 1 indicates full overlap (the distributions are the same).  The index is estimated using samples from the two distributions using the ``overlapping" package in R \citep{Pastore2019}.  We compute the overlapping index for the DIC values from the true model and compare these values to the DIC values from the similar and wrong model.  Ideally, the overlapping index would be 0 for both of these comparison models.  

Table \ref{tab:overlap} shows the overlapping index for the posterior sample sizes considered.  For even 2,000 saved posterior draws, the DIC values between the true model and the wrong model are easy to distinguish, with the overlapping index between the two equal to 0.  However, the DIC values overlap quite a bit between the true and similar models, with an overlapping index of 0.5341.  At 20,000 saved posterior saved values, the overlapping index reduces to 0.0022.  

\begin{table}
    \centering
        \caption{Overlap index comparing distribution of the computed DIC values from the true model to similar model (2nd column) and the wrong model (3rd column) for different posterior sample sizes.}
    \label{tab:overlap}
    \begin{tabular}{c|cc}
    \hline
    & \multicolumn{2}{c}{Overlap Index}\\
    Sample Size & True-Similar & True-Wrong \\\hline\hline
       2,000  & 0.5341 & 0 \\
       8,000  & 0.0405 & 0 \\
       16,000 & 0.0861 & 0 \\
       20,000 & 0.0022 & 0 \\\hline
    \end{tabular}

\end{table}

We also examined the ESS values and the posterior standard deviations of these three models to examine the Monte Carlo variability for these values.  The smallest ESS for the parameters of the 20 times we fit the true model with 20,000 saved posterior draws was 309, or more than 1.5\% of the sample size. The largest posterior standard deviation of the change points was 1154.  We use these as guidelines for the convergence cut-offs for the temperature data analysis.

\subsection{Utah Temperature Data}

We now explore the forward selection process on the Utah minimum daily temperature measurements.  In contrast to the simulation study, we do not expect our change point model to exactly match the true data generating process.  Therefore, our goal is not to identify the true model, but rather a good model that we can explore to better understand temperature behavior changes at different locations.  

For the forward selection process, we use the same prior distributions that we used for the synthetic data.  We fit the MCMC using a burn-in of 20,000 iterations and thin every 5th.  At each step of the forward selection process, a model is considered if the ESS of all parameters are at least 1\% of the saved iterations and the standard deviation of the change points are smaller than 2500. 

We fit the forward selection process three different times and all three arrived at different, but similar, final models.  When the selection processes diverged, it was either because a similar model had a smaller DIC or because one of the models met the ESS criteria in only one run of the selection processes.  Based on these results, we chose five models to compare: the three chosen based on the selection processes and two additional models that had smaller DIC values but did not meet the initial ESS criteria.  We fit these five models for 40,000 total saved iterations (burning 22,000 and thinning by 5).  We examined the Geweke convergence diagnostic \citep{Geweke1992} for all parameters for each model using the ``coda" package \citep{coda2006}.  All diagnostic values fell between -3 and 3.  Table \ref{tab:finalmods} shows the number of change points at each location, DIC, MSE, and smallest ESS for each parameter of each model.

\begin{table}[]
    \centering
        \caption{Number of change points for each location, DIC, MSE, and smallest ESS for the five fitted models chosen from the forward selection process for the Utah temperature data.}
    \label{tab:finalmods}
   \hspace*{-.2in} \begin{tabular}{c|cccccccc|ccc}
    \hline
    & \multicolumn{8}{c|}{Location} & \\
        &1 & 2&3&4&5&6&7&8 &  &&\\
Model  & {\scriptsize Blanding} &  {\scriptsize Bluff} &  {\scriptsize Cedar} &  {\scriptsize Escalante} &  {\scriptsize Fillmore} &  {\scriptsize SLC} &  {\scriptsize Woodruff} &  {\scriptsize Zion} & DIC & MSE & Min ESS \\\hline\hline
1 &    1  &  2  &  2  &  0 &   1  &  1  &  1  &  0 & 1036232 & 18.524 & 630.15\\
2 &    2  &  2  &  2  &  0  &  1  & 1  &  1   & 0 & 1036102 & 18.510 & 331.42\\
3 &    2  &  2  &  2   & 0 &   1  &  2 &   1   &0 & 1036122 & 18.512 & 309.43\\
4 &    2  &  1  &  2  &  0  &  1   & 2  &  1   & 0& 1036306 & 18.531 & 815.90\\
5 &    2  &  2   & 1    &0 &   1    &2   & 1    &0& 1036238& 18.524 & 509.78\\\hline
    \end{tabular}
\end{table}

\begin{figure}
     \centering
     \begin{subfigure}[b]{0.45\textwidth}
         \centering
         \includegraphics[width=\textwidth]{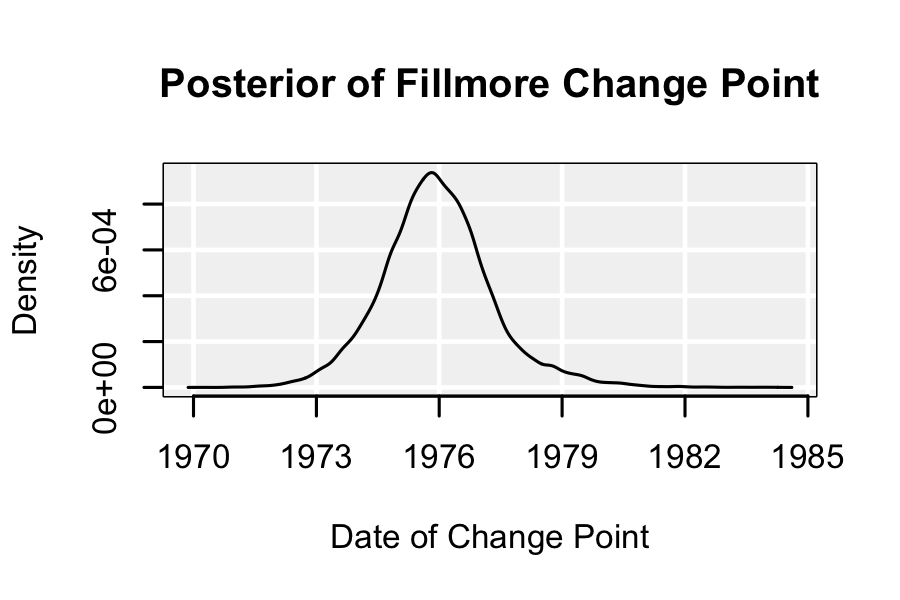}
         \caption{}
         \label{fig:Fchange}
     \end{subfigure}
     \hfill
     \begin{subfigure}[b]{0.45\textwidth}
         \centering
         \includegraphics[width=\textwidth]{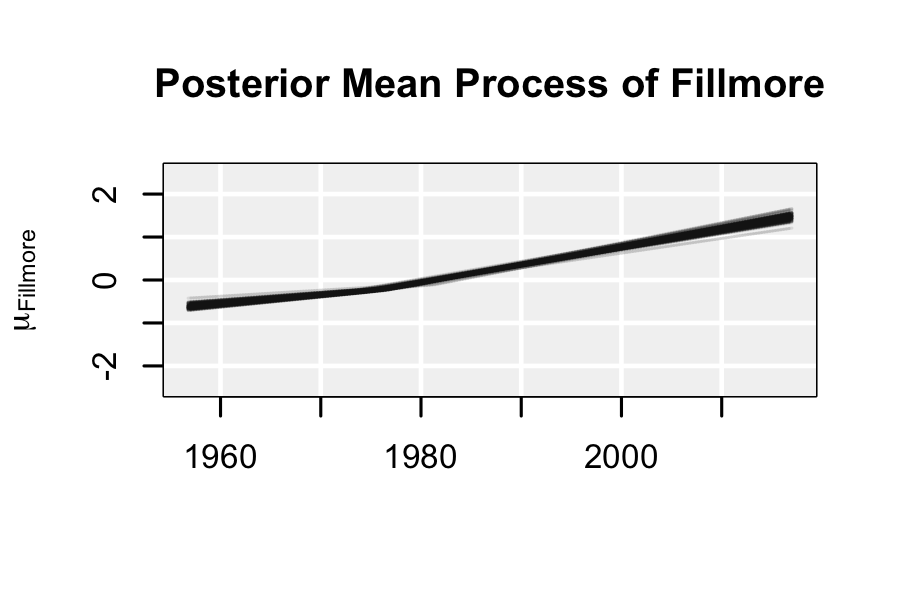}
         \caption{}
         \label{fig:Fmean}
     \end{subfigure}
     \hfill
     \begin{subfigure}[b]{0.45\textwidth}
         \centering
         \includegraphics[width=\textwidth]{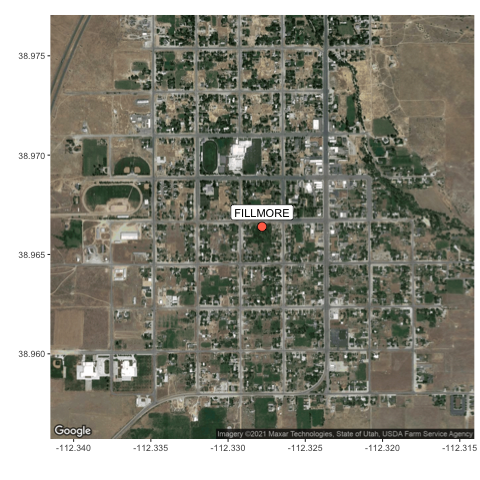}
         \caption{}
         \label{fig:Fcity}
     \end{subfigure}
      \hfill
     \begin{subfigure}[b]{0.45\textwidth}
         \centering
         \includegraphics[width=\textwidth]{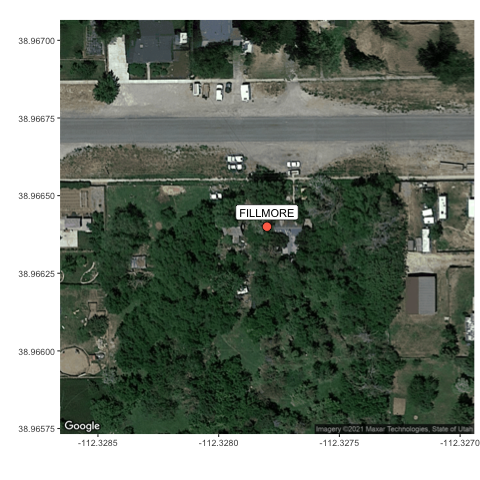}
         \caption{}
         \label{fig:FStation}
     \end{subfigure}
        \caption{(a) Posterior distribution for the date of Fillmore's change point, $\tau_{1,5}$. (b) 100 realizations from the posterior of the mean process of Fillmore, $\bfmu_5$.  (c) Satellite view of the city of Fillmore.  The station is marked with the orange dot.  (d) Satellite view of the location of the station, marked in orange. }
        \label{fig:Fillmore}
\end{figure}

All five models are consistent in identifying which locations do not have change points: Escalante and Zion National Park.  Both of these locations are part of national preserves (monument and park, respectively), thus, urbanization is unlikely.  All five models likewise find that both Fillmore and Woodruff have one change point.  Considering Fillmore, all five models identify a change point that ranges between 1970 and 1980. Figure \ref{fig:Fchange} shows the posterior distribution of the change point and the estimated mean process of Fillmore for Model 1.  Figure \ref{fig:Fmean} shows 100 realizations from the posterior of the mean process for Fillmore, $\bfmu_5$, where a change point in the mid-to-late 70's is visible.  Although Fillmore is a relatively small city (current population around 2000), during this time period, the population of Fillmore increased by 50\% \citep{census1980}.  The station itself is in the middle of the city \citep[see Figures \ref{fig:Fcity} and \ref{fig:FStation}, mapped using R package `ggmap;'][]{ggmap}, thus, temperatures at that location are likely sensitive to city-wide growth.  Woodruff is a much smaller city and, to save on space, a discussion of the change point identified by the five models for this location is provided in the Supplementary Document.

\begin{figure}
    \centering
    \includegraphics[width=6.5in]{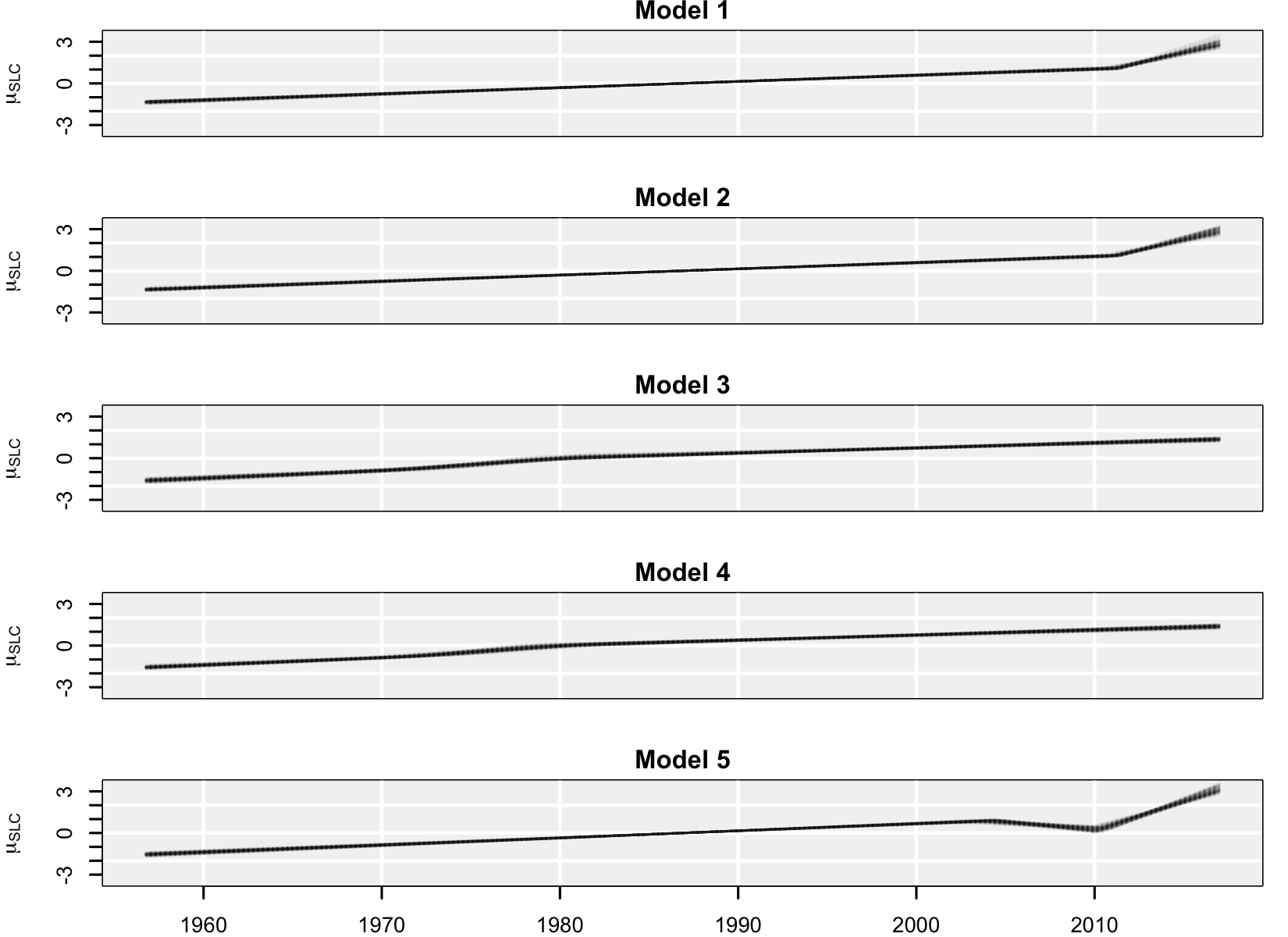}
    \caption{50 realizations from the posterior distribution on the mean process of the Salt Lake City Airport station, $\bfmu_6$, for each of the fitted models listed in Table \ref{tab:finalmods}.}
    \label{fig:SLCmean}
\end{figure}

The five models disagree on which locations should have two change points.  This is not surprising since, for example, having more change points in the model creates more dependence between model parameters and makes it more difficult for the MCMC to converge.  Consider the Salt Lake City Airport station, where two models fit one change point and the other three models fit two change points.  In this case, the mean processes are quite diverse.  Figure \ref{fig:SLCmean} shows 50 realizations from the posterior distribution of the mean process for each of the five fitted models.  Models 1, 2, and 5 estimated increasing mean temperatures at the end of the time period around 2010, while Models 3 and 4 estimated changes in the mean process around the 1970's and 1980's. The Salt Lake International Airport has undergone several large improvements over the last sixty years.  According to the airport's website \citep{SLC2020}, ``from 1975 to 1980, the airport grew to 7500 acres," and two terminals were expanded and remodeled in the early 1980's.  Additionally, construction began on a complete overhaul of the airport terminals and parking structures in 2014.  It may be that only two change points is not sufficient to model the construction changing land surface happening around this station.

Similar figures and discussions are provided for all eight locations in the Supplementary Document.  

\section{Discussion}\label{sec:discussion}

In this work we proposed a change point selection process as a practical approach to determining an appropriate change point model for spatio-temporal data. This is useful in the current application examining changes in temperature increases and decreases over time. Other uses for this modeling approach include other environmental or economic applications.  The data analysis identified change points that made sense with known urbanization patterns of some locations.  

The forward selection process does not necessarily evaluate the same models when fit multiple times. This is because the selection processes explored different combinations of change points and thus there arose inconsistencies in the chosen models from each repetition. This highlights the need for the change point selection methodology to fully explore the space of change point combinations. Other methods that potentially do this, such as reversible jump MCMC \citep{Green1995}, often have difficult implementation and similar difficulties exploring the parameter space. As demonstrated using the synthetic data set, the methodology we propose allows the selection process to explore as many change point combinations as possible through a simpler implementation.


Future work will examine these same models but for maximum daily temperature measurements and the difference between the maximum and minimum daily temperature measurements and compare change points identified from these alternative temperature measures.



\bibliography{refs}

\begin{thebibliography}{49}
\newcommand{\enquote}[1]{``#1''}
\expandafter\ifx\csname natexlab\endcsname\relax\def\natexlab#1{#1}\fi

\bibitem[\protect\citename{Akbari and Konopacki, }2005]{Akbari2005}
Akbari, H. and Konopacki, S. (2005).
\newblock \enquote{Calculating energy-saving potentials of heat-island
  reduction strategies.}
\newblock {\em Energy Policy\/}, 33, 721--756.

\bibitem[\protect\citename{Arthur, }2018]{Arthur2018}
Arthur, D.~B. (2018).
\newblock \enquote{A Bayesian Spatio-Temporal Change Point Model for
  Identifying Urban Heat Islands in Utah.}
\newblock Master's thesis, Brigham Young University, Provo, UT.

\bibitem[\protect\citename{Balling~Jr and Brazel, }1988]{Balling1988}
Balling~Jr, R. and Brazel, S. (1988).
\newblock \enquote{High-resolution surface temperature patters in a complex
  urban terrain.}
\newblock {\em Photogrammetric Engineering \& Remote Sensing\/}, 54,
  1289--1293.

\bibitem[\protect\citename{Benson and Friel, }2018]{Benson2018}
Benson, A. and Friel, N. (2018).
\newblock \enquote{Adaptive MCMC for multiple changepoint analysis with
  applications to large datasets.}
\newblock {\em Electronic Journal of Statistics\/}, 12, 3365--3396.

\bibitem[\protect\citename{Calder et~al., }2002]{Calder2002}
Calder, C.~A., Holloman, C., and Higdon, D. (2002).
\newblock \enquote{Exploring Space-Time Structure in Ozone Concentration Using
  a Dynamic Process Convolution Model.}
\newblock {\em Lecture Notes in Statistics\/}, 167.

\bibitem[\protect\citename{Changnon et~al., }1996]{Changnon1996}
Changnon, S.~A., Kunkel, K.~E., and Reinke, B.~C. (1996).
\newblock \enquote{Impacts and Responses to the 1995 Heat Wave: A Call to
  Action.}
\newblock {\em Bulletin of the American Meteorological Society\/}, 77,
  1497--1506.

\bibitem[\protect\citename{Chauveau and Vandekerkhove, }2002]{Chauveau2002}
Chauveau, D. and Vandekerkhove, P. (2002).
\newblock \enquote{Improving Convergence of the Hastings- Metropolis Algorithm
  with an Adaptive Proposal.}
\newblock {\em Scandinavian Journal of Statistics\/}, 29, 13--29.

\bibitem[\protect\citename{Congdon, }2005]{Congdon2005}
Congdon, P. (2005).
\newblock \enquote{Bayesian predictive model comparison via parallel sampling.}
\newblock {\em Computational Statistics \& Data Analysis\/}, 48, 735--753.

\bibitem[\protect\citename{Cressie and Johanneson, }2008]{Cressie2008}
Cressie, N. and Johanneson, G. (2008).
\newblock \enquote{Fixed rank kriging for very large spatial data sets.}
\newblock {\em Journal of the Royal Statistical Society\/}, 70, 209--226.

\bibitem[\protect\citename{Fernhead, }2006]{Fernhead2006}
Fernhead, P. (2006).
\newblock \enquote{Exact and efficient Bayesian inference or multiple
  changepoint problems.}
\newblock {\em Statistical Computing\/}, 16, 203--213.

\bibitem[\protect\citename{Flegal et~al., }2020]{Flegal2020}
Flegal, J.~M., Hughes, J., Vats, D., and Dai, N. (2020).
\newblock {\em mcmcse: Monte Carlo Standard Errors for MCMC\/}.
\newblock Riverside, CA, Denver, CO, Coventry, UK, and Minneapolis, MN.
\newblock R package version 1.4-1.

\bibitem[\protect\citename{Furgale et~al., }2012]{Furgale2012}
Furgale, P., Barfoot, T.~D., and Sibley, G. (2012).
\newblock \enquote{Continuous-time batch estimation using temporal basis
  functions.}
\newblock In {\em 2012 IEEE International Conference on Robotics and
  Automation\/},  2088--2095.

\bibitem[\protect\citename{Geweke, }1992]{Geweke1992}
Geweke, J. (1992).
\newblock \enquote{Evaluating the accuracy of sampling-based approaches to
  calculating posterior moments.}
\newblock {\em Bayesian Statistics\/}, 4, 169--193.

\bibitem[\protect\citename{Gong and Flegal, }2016]{Gong2016}
Gong, L. and Flegal, J.~M. (2016).
\newblock \enquote{A Practical Sequential Stopping Rule for High-Dimensional
  Markov Chain Monte Carlo.}
\newblock {\em Journal of Computational and Graphical Statistics\/}, 25,
  684--700.

\bibitem[\protect\citename{Green, }1995]{Green1995}
Green, P. (1995).
\newblock \enquote{Reversible jump Markov chain Monte Carlo computation and
  Bayesian model determination.}
\newblock {\em Biometrika\/}, 82, 711–732.

\bibitem[\protect\citename{Gregory, }2010]{Gregory2010}
Gregory, P.~C. (2010).
\newblock \enquote{Bayesian exoplanet tests of a new method for MCMC sampling
  in highly correlated model parameter spaces.}
\newblock {\em Monthly Notices of the Royal Astronomical Society\/}, 410,
  94--110.

\bibitem[\protect\citename{Grimmond, }2007]{Grimmond2007}
Grimmond, S. (2007).
\newblock \enquote{Urbanization and global environmental change: local effects
  of urban warming.}
\newblock {\em The Geographical Journal\/}, 173, 1, 83--88.

\bibitem[\protect\citename{Haario et~al., }2001]{Haario2001}
Haario, H., Saksman, E., and Tamminen, J. (2001).
\newblock \enquote{An adaptive Metropolis algorithm.}
\newblock {\em Bernoulli\/}, 7, 223--242.

\bibitem[\protect\citename{Haario et~al., }2005]{Haario2005}
--- (2005).
\newblock \enquote{Componentwise adaptation for high dimensional MCMC.}
\newblock {\em Computational Statistics\/}, 20, 265--273.

\bibitem[\protect\citename{Hartz et~al., }2006]{Hartz2006}
Hartz, D., Prashad, L., Hedquist, B., Golden, J., and Brazel, A. (2006).
\newblock \enquote{Linking satellite images and hand-held infrared thermography
  to observed neighborhood climate conditions.}
\newblock {\em Remote Sensing of Environment\/}, 104, 190--200.

\bibitem[\protect\citename{Higdon, }2002]{Higdon2002}
Higdon, D. (2002).
\newblock \enquote{Space and Space-Time Modeling using Process Convolutions.}
\newblock {\em Quantitative Methods for Current Environmental Issues\/},
  37–56.

\bibitem[\protect\citename{Hoffman et~al., }2012]{Hoffman2012}
Hoffman, P., Krueger, O., and Schlunzen, K.~H. (2012).
\newblock \enquote{A statistical model for the urban heat island and its
  application to a climate change scenario.}
\newblock {\em International Journal of Climatology\/}, 32, 1238--1248.

\bibitem[\protect\citename{IPCC, }2019]{IPCC2019}
IPCC (2019).
\newblock {\em Climate Change and Land: an IPCC special report on climate
  change desertification, land degradation, sustainable land management, food
  security, and greenhouse gas fluxes in terrestrial ecosystems\/}.
\newblock P.R. Shukla, J. Skea, E. Calvo Buendia, V. Masson-Delmotte, H.-O.
  Pörtner, D. C. Roberts, P. Zhai, R. Slade, S. Connors, R. van Diemen, M.
  Ferrat, E. Haughey, S. Luz, S. Neogi, M. Pathak, J. Petzold, J. Portugal
  Pereira, P. Vyas, E. Huntley, K. Kissick, M. Belkacemi, J. Malley, (eds.).

\bibitem[\protect\citename{Kahle and Wickham, }2013]{ggmap}
Kahle, D. and Wickham, H. (2013).
\newblock \enquote{ggmap: Spatial Visualization with ggplot2.}
\newblock {\em The R Journal\/}, 5, 1, 144--161.

\bibitem[\protect\citename{Katzfuss and Cressie, }2012]{Katzfuss2012}
Katzfuss, M. and Cressie, N. (2012).
\newblock \enquote{Bayesian hierarchical spatiotemporal smoothing for very
  large datasets.}
\newblock {\em Environmetrics\/}, 23, 94--107.

\bibitem[\protect\citename{Kim and Siegmund, }1989]{Kim1989}
Kim, H.-J. and Siegmund, D. (1989).
\newblock \enquote{The likelihood ratio test for a change-point in simple
  linear regression.}
\newblock {\em Biometrika\/}, 76, 409--423.

\bibitem[\protect\citename{Konidaris and Osentoski, }2008]{Konidaris2008}
Konidaris, G. and Osentoski, S. (2008).
\newblock \enquote{Value function approximation in reinforcement learning using
  the Fourier basis.}
\newblock {\em Computer Science Department Faculty Publication Series\/}, 101.

\bibitem[\protect\citename{Lee and M., }2011]{Lee2011}
Lee, D.-J. and M., D. (2011).
\newblock \enquote{P-spline ANOVA-type interaction models for spatio- temporal
  smoothing.}
\newblock {\em Statistical Modelling\/}, 11, 49--69.

\bibitem[\protect\citename{Majumdar et~al., }2005]{Majumdar2005}
Majumdar, A., Gelfand, A., and Banerjee, S. (2005).
\newblock \enquote{Spatio-temporal change-point modeling.}
\newblock {\em Journal of Statistical Planning and Inference\/}, 130, 149--166.

\bibitem[\protect\citename{Mortensen et~al., }2018]{Heaton2018}
Mortensen, J., Heaton, M.~J., and Wilhelmi, O.~V. (2018).
\newblock \enquote{Urban Heat Risk Mapping of Houston, Texas using Multiple
  Point Patterns.}
\newblock {\em Journal of the Royal Statistical Society, Series C\/}, 67, 1,
  83--102.

\bibitem[\protect\citename{Navarra and Simoncini, }2010]{Navarra2010}
Navarra, A. and Simoncini, V. (2010).
\newblock {\em A guide to empirical orthogonal functions for climate data
  analysis\/}.
\newblock Springer Science \& Business Media.

\bibitem[\protect\citename{Novosad and Reader, }2016]{Novosad2016}
Novosad, P. and Reader, A.~J. (2016).
\newblock \enquote{MR-guided dynamic PET reconstruction with the kernel method
  and spectral temporal basis functions.}
\newblock {\em Physics in Medicine \& Biology\/}, 61.

\bibitem[\protect\citename{Oke, }1973]{oke1973}
Oke, T. (1973).
\newblock \enquote{City size and the urban heat island.}
\newblock {\em Atmospheric Environment\/}, 7, 769--779.

\bibitem[\protect\citename{Pastore and Calcagni, }2019]{Pastore2019}
Pastore, M. and Calcagni, A. (2019).
\newblock \enquote{Measuring Distribution Similarities Between Samples: A
  Distribution-Free Overlapping Index.}
\newblock {\em Frontiers in Psychology\/}, 10, 1089.

\bibitem[\protect\citename{Plummer et~al., }2006]{coda2006}
Plummer, M., Best, N., Cowles, K., and Vines, K. (2006).
\newblock \enquote{CODA: Convergence Diagnosis and Output Analysis for MCMC.}
\newblock {\em R News\/}, 6, 1, 7--11.

\bibitem[\protect\citename{{R Core Team}, }2021]{R2021}
{R Core Team} (2021).
\newblock {\em R: A Language and Environment for Statistical Computing\/}.
\newblock R Foundation for Statistical Computing, Vienna, Austria.

\bibitem[\protect\citename{Roberts and Rosenthal, }2009]{Roberts2009}
Roberts, G.~O. and Rosenthal, J.~S. (2009).
\newblock \enquote{Examples of Adaptive MCMC.}
\newblock {\em Journal of Computational and Graphical Statistics\/}, 18,
  349--367.

\bibitem[\protect\citename{Saaroni et~al., }2000]{Saaroni2000}
Saaroni, H., Ben-Dor, E., Bitan, A., and Potchter, O. (2000).
\newblock \enquote{Spatial distribution and microscale characteristics of the
  urban heat island in Tel-Aviv, Israel.}
\newblock {\em Landscape and Urban Planning\/}, 48, 1--18.

\bibitem[\protect\citename{{Salt Lake City International Aiport},
  }2020]{SLC2020}
{Salt Lake City International Aiport} (2020).
\newblock Accessed: 2021-05-21.

\bibitem[\protect\citename{Seber and Lee, }2003]{Seber2003}
Seber, G. A.~F. and Lee, A.~J. (2003).
\newblock {\em Linear Regression Analysis\/}.
\newblock 2nd ed. John Wiley \& Sons, Inc.

\bibitem[\protect\citename{Sloan et~al., }2017]{Sloan2017}
Sloan, C., Heaton, M., Kang, S., Berrett, C., Gebretsadik, T., Sicignano, N.,
  Evans, A., Lee, R., and Hartert, T. (2017).
\newblock \enquote{The Impact of Temperature and Relative Humidity on
  Spatiotemporal Patterns of Infant Bronchiolitis Epidemics in the Contiguous
  United States.}
\newblock {\em Health \& Place\/}, 45, 46--54.

\bibitem[\protect\citename{Spiegelhalter et~al., }2002]{Spiegelhalter2002}
Spiegelhalter, D.~J., Best, N.~G., Carlin, B.~P., and Van Der~Linde, A. (2002).
\newblock \enquote{Bayesian measures of model complexity and fit.}
\newblock {\em Journal of the Royal Statistical Society\/}, 64, 583--639.

\bibitem[\protect\citename{Stabler et~al., }2005]{stabler2005}
Stabler, L.~B., Martin, C.~A., and Brazel, A.~J. (2005).
\newblock \enquote{Microclimates in a desert city were related to land use and
  vegetation index.}
\newblock {\em Urban Forestry \& Urban Greening\/}, 3, 137--147.

\bibitem[\protect\citename{Stein and Weiss, }1971]{Stein1971}
Stein, E.~M. and Weiss, G. (1971).
\newblock {\em Introduction to Fourier Analysis on Euclidean Spaces
  (PMS-32)\/}.
\newblock Princeton University Press.

\bibitem[\protect\citename{{United States Census Bureau}, }2020]{census1980}
{United States Census Bureau} (2020).
\newblock \enquote{Census of Population and Housing: 1980 Census of Population,
  Utah.}
\newblock Accessed: 2021-05-21.

\bibitem[\protect\citename{USUCC, }2017]{USUCC2017}
USUCC (2017).
\newblock \enquote{Utah~State~University~Climate~Center: Utah Climate Center
  Database.}
\newblock Accessed: 2020-11-24.

\bibitem[\protect\citename{Weng et~al., }2004]{Weng2004}
Weng, Q., Lu, D., and Schubring, J. (2004).
\newblock \enquote{Estimation of land surface temperature–vegetation
  abundance relationship for urban heat island studies.}
\newblock {\em Remote Sensing of Environment\/}, 89, 467--483.

\bibitem[\protect\citename{Yu et~al., }2008]{Yu2008}
Yu, Q., Scribner, R., Carlin, B., Theall, K., Simonsen, N., Ghosh-Dastidar, B.,
  Cohen, D., and Mason, K. (2008).
\newblock \enquote{Multilevel spatio-temporal dual changepoint models for
  relating alcohol outlet destruction and changes in neighbourhood rates of
  assaultive violence.}
\newblock {\em Geospatial Health\/}, 2.

\bibitem[\protect\citename{Zumer et~al., }2008]{Zumer2008}
Zumer, J.~M., Attias, H.~T., Sekihara, K., and Nagarajan, S.~S. (2008).
\newblock \enquote{Probabilistic algorithms for MEG/EEG source reconstruction
  using temporal basis functions learned from data.}
\newblock {\em NeuroImage\/}, 41, 924--940.

\end{thebibliography}

\end{document}